# B Physics From NRQCD With Dynamical Fermions[*]

Presented by S. Collins[a] [†]

[a]SCRI, Florida State University, Tallahassee, Fl 32306-4052, USA

We present *preliminary* results for the spectrum and decay constants of B mesons using NRQCD heavy and Wilson light quarks on configurations at $\beta = 5.6$ with two dynamical flavours of staggered fermions. All terms to order $1/M_Q$ are included in the NRQCD action; matrix elements are corrected to this order by calculating the small components of the heavy quark propagator.

## 1. Introduction

With the viability of using NRQCD to simulate heavy quarks established in quarkonium systems [1] there is increasing interest in applying this approach to heavy-light mesons. Here, the heavy quark mass plays a more minor role, entering the spectrum, for example, only at the level of the hyperfine splitting. Initially, therefore, we consider terms up to $O(1/M_Q^0)$ in the NRQCD lagrangian:

$$L_{\text{NRQCD}} = Q^\dagger \left\{ -D_t + \frac{D^2}{2M_Q^0} + \frac{g}{2M_Q^0}\sigma \cdot B \right\} Q \quad (1)$$

where $M_Q^0$ is the bare heavy quark mass. Tadpole improvement is used throughout. In order to obtain matrix elements to this level of accuracy, the small components of the heavy quark propagator must be reconstructed [2]. Of particular interest is the axial-vector current; the relation between the current in full QCD and that calculated in NRQCD is given by

$$\langle \bar{q}\gamma_5\gamma_0 q_h|PS\rangle_{QCD} = \langle q^\dagger Q|PS\rangle_{\text{NRQCD}} +$$
$$\frac{1}{2M_Q^0}\langle q^\dagger(-i\sigma\cdot D)Q|PS\rangle_{\text{NRQCD}} \quad (2)$$

at $O(1/M_Q^0)$. The $M_Q$ dependence of the current can be made more explicit by relating the NRQCD current to that in the infinite mass limit. In addition to (2), both the kinetic and hyperfine terms in the action contribute $O(1/M_Q^0)$ corrections to $\langle q^\dagger Q|PS\rangle_\infty$ through $|PS\rangle_{\text{NRQCD}}$.

---

[*]Talk presented at Lattice '94, September 27–October 1, 1994, Bielefeld, Germany
[†]In collaboration with A. Ali Khan and C. T. H. Davies, Glasgow Univ.; J. Shigemitsu, Ohio State Univ.; U. M. Heller and J. H. Sloan, SCRI at Florida State Univ.

Table 1
† from NRQCD.

|  | Force | $M_\rho$ | $M_N$ | $\Upsilon$† |
|---|---|---|---|---|
| $a^{-1}$ (GeV) | 2.06 (9) | 2.14 (6) | 1.8 (1) | 2.4 (1) |

## 2. Simulation Details

The simulation was performed on 100 $16^3 \times 32$ lattices at $\beta = 5.6$ with two flavours of dynamical staggered fermions, generated by the HEMCGC collaboration [3]. The light quark propagators were generated using the Wilson action, without an O(a) improvement term, at two values for the light quark ($\kappa_l = 0.1585, 0.1600$). The heavy quark propagators were computed at six values of $M_Q^0$ ranging from 0.8 to 4.0; the propagator in the static limit was also calculated. Further details of our method and analysis are given in [2].

## 3. Lattice Spacing and Systematic Errors

Table 1 details the inverse lattice spacings for these lattices. Note that there is a large spread in $a^{-1}$ between quantities at dissimilar scales, for example the force and $\Upsilon$ spectrum; with $n_f = 2$ dynamical fermions, incorrect running of the strong coupling is unlikely to account for this discrepancy. In addition, table 1 shows that there is significant disagreement between $a^{-1}$'s from light spectroscopy (much larger than the statistical errors of $\approx 5\%$). This suggests that there are large systematic errors in our light quark propagators, probably arising from the $O(ma)$ error in the Wilson action and/or finite volume effects. Thus, only very rough predictions are possible and we shall concentrate on the $M_Q$ dependence of physical quantities. The important systematic error here is the $O(1/M_Q)$ error in $1/M_Q$ terms, from the truncation of the NRQCD series. Naively this is $\approx 10\%$ for the B meson. $O(g^2)$ corrections to the coefficient of the hyperfine term (1 at tree-level) are expected to be of a similar magnitude.

## 4. Mass Splittings

In HQET the mass of a heavy-light meson $M(Ql)$ at $O(1/M_Q)$ is given by [4]

$$M(Ql) = M_Q + \epsilon_l + \frac{\Delta_l}{M_Q} \langle S_Q \cdot S_l \rangle \qquad (3)$$

where $\epsilon_l$ is the binding energy of the light degrees of freedom in the static approximation and $\Delta_l \propto |\psi(0)|$. At this order, eqn. (3) leads to a simple prediction for the hyperfine mass splitting

$$M_V(Ql) - M_{PS}(Ql) \propto 1/M_Q, \qquad (4)$$

which is roughly supported by experiment: $(B^* - B)/(D^* - D) \approx M_D/M_B \approx 3$. Figure 1 presents the results for the hyperfine splitting as a function of $1/M_{PS}(Ql)$ for $\kappa_l = 0.1585$. The uncorrelated linear fit shown suggests that the results are in rough agreement with the prediction. In the region of the B meson the $O(1/M_Q)$ systematic errors in the splitting are of the same order of magnitude as the statistical errors. No significant dependence of the hyperfine splitting on the light quark mass was found, as seen in experiment. Due to the $\sim 1/M_Q$ dependence, the splitting is doubly sensitive to the uncertainty in $a^{-1}$, and we quote a large range, $20 - 40$ MeV ($a^{-1} = 1.8 - 2.4$); slightly below experiment, $B^* - B = 46(1)$MeV.

Taking the spin average of $M(Ql)$:

$$\bar{M}(Ql) \equiv \frac{1}{4}(3M_V(Ql) + M_{PS}(Ql)) = M_Q + \epsilon_l. \quad (5)$$

This suggests that the quantity $\bar{M}(Qs) - \bar{M}(Qd)$ should be independent of $M_Q$, where $M(Qs)$ is the mass of a meson containing a strange quark. Experimentally, this splitting is equal to 101(2) MeV for the D meson compared to 89(5) MeV for the B meson. The results for $\bar{M}(Qs) - \bar{M}(Qd)$ in lattice units (l.u.), for a range of values of $M_Q^0$ are shown in table 2, where $M_\phi/M_\rho$ was used to obtain $\kappa_s$. The splitting is fairly independent of the heavy quark mass, as expected, although there may be a small decrease with increasing $M_Q^0$. In physical units the splitting is between $100 - 130$ MeV, slightly higher than experiment.

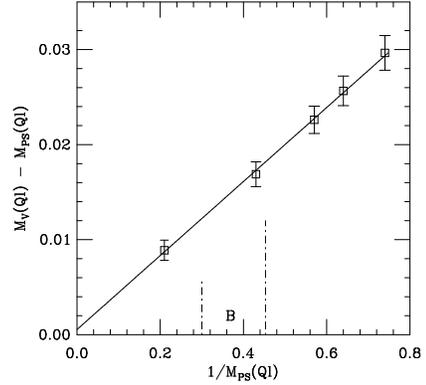

Figure 1. The hyperfine splitting vs $1/M_{PS}(Ql)$ in l.u. for $\kappa_l = 0.1585$. The dashed lines indicate $1/M_B$ in l.u. using $a^{-1} = 1.8$ GeV and 2.4 GeV.

Table 2

| $M_Q^0$ | $\infty$ | 1.7 | 1.2 | 0.8 |
|---|---|---|---|---|
| $\bar{M}(Qs) - \bar{M}(Qd)$ | 61(9) | 61(5) | 63(5) | 65(5) |
| $\frac{f_{PS}(Qs)}{f_{PS}(Qd)}$ | | 1.23 (4) | 1.25 (5) | 1.26 (2) | 1.26 (2) |

## 5. Pseudoscalar Decay Constant

The amplitude of $\langle q^\dagger Q | PS \rangle_\infty$ is independent of $M_Q$ and this gives rise to the scaling law:

$$f_{PS} \sqrt{M_{PS}(Ql)} = \text{const.}(\alpha_s(M_{PS}(Ql)))^{-2/\beta_0}. \quad (6)$$

Although naively all $O(1/M_Q)$ corrections to eqn. (6) are of the same order of magnitude ($\sim 10\%$ at the B meson), the correction from the kinetic energy term should dominate since the spin dependent terms in (1) and (2) are further suppressed as they violate spin symmetry.

Figure 2 shows the quantity

$$\Phi(M(Ql)) \equiv Z_A^{-1} f_{PS} \sqrt{M_{PS}(Ql)}$$
$$(\alpha_s(M_{PS}(Ql))/\alpha_s(M_B))^{2/\beta_0} \quad (7)$$

as a function of $1/M_{PS}(Ql)$ for $\kappa_l = 0.1585$, both with and without the matrix element correction. The quadratic fits shown are uncorrelated, and the dashed line indicates (roughly) the extrapolation of the results to $M_Q^0 = \infty$. The matrix element correction reduces $\Phi(M(Ql))$ by $10-15\%$ of $\Phi(\infty)$ in the region of the B meson, compared to a $15-20\%$ reduction arising from the kinetic and hyperfine terms in the action. Taking the spin-average of $f_{PS}$ and $f_V$, without the matrix element correction, would separate these two contributions and show whether all three terms are of a similar magnitude. $O(1/M_Q)$ corrections to (6) of $25-35\%$ at the B meson are much larger than naively expected; this suggests the $O(1/M_Q^2)$ corrections may be as large as $10\%$ ($> 3\%$ statistical errors), rather than the $1\%$ expected. In addition, $\Phi(M(Ql))$ appears to be non-linear in $1/M_{PS}(Ql)$ and this is a further indication that it is necessary to go to the next order in $1/M_Q$, particularly when connecting with Wilson fermion results close to the D meson.

Note that $Z_A$, also $M_Q$ dependent, must be included in order to properly extract the $1/M_Q$ corrections to (6); we are in the process of calculating $Z_A$. However, the static result, also shown in figure 2, should agree with the extrapolation of the finite $M_Q^0$ results, since it is the large mass limit of NRQCD. The large discrepancy seen is probably largely due to significant contributions from excited states remaining in the static result; with noise dominating at timeslice 12 (compared to $\sim 25$ for the finite values of $M_Q^0$ used) a multi-smearing multi-exponential analysis is necessary. Nevertheless, using $Z_A^{stat} \sim 0.67$ [5], we obtain $f_B^{stat} = 200 - 310$ MeV.

A quantity which does not depend on $Z_A$, and which is only weakly dependent on $a^{-1}$ ($\sim 10\%$) through $\kappa_s$, is the ratio $f_{PS}(Qs)/f_{PS}(Qd)$ presented in table 2 (the matrix element correction is included but it has only a $1\%$ effect). The ratio is insensitive to $M_Q^0$, and so independent of the uncertainties in the B meson mass. The results are in agreement with previous calculations,

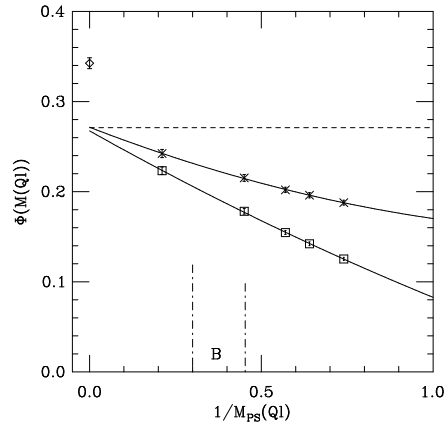

Figure 2. $\Phi(M_{PS}(Ql))$ vs $1/M_{PS}(QL))$ in l.u. for $\kappa_l = 0.1585$, including (squares) and omitting (crosses) the matrix element contribution; the static result (diamond) is also shown.

which have found $f_{PS}(Qs)/f_{PS}(Qd) \approx 1.1 - 1.3$.

## 6. Conclusions

The preliminary results look very promising, particularly for $f_{PS}$ where the $1/M_Q$ corrections to the scaling law can be computed separately and accurately; there is an indication that it may be necessary to go to second order in $1/M_Q$. It is essential to improve the light quark propagators, and the determination of $a^{-1}$, through the use of an $O(a)$-improved fermion action.

S. Collins would like to thank her collaborators for many useful discussions. This work was supported by the DOE; the computations were carried out at SCRI and in part at the Ohio Supercomputer Center.